\documentstyle[aps,prl,multicol,epsfig,psfig,epsf]{revtex}
\begin{document}
\draft
\title{Competition between spin glass order and strong coupling
superconductivity in a single - species fermion model}
\author{H. Feldmann and R. Oppermann
\\
}\address{
Institut f\"ur Theoretische Physik,\\ Univ. W\"urzburg,
D--97074 W\"urzburg, FRG}
\date{\today}
\maketitle
\begin{abstract}
The phase diagram of a single species fermion model allowing for local 
pairing superconductivity (SC) and spin glass order (SG) is derived as a 
function of chemical potential $\mu$ and ratio $r\equiv v/J$ between 
attractive coupling $v$ and frustrated magnetic interaction J. 
For ratios larger than a characteristic $r_c(\mu)$, superconductivity does
not allow for SG order, while for smaller values a very detailed phase
diagram arises with entangled spin glass and superconducting transitions.
Our results for the Green's functions show that superconductivity
occurring in the magnetic interaction band is of gapless type with a crossover
from strongly gapless, within a certain range below $T_c$, to very weakly
gapless in a wide low temperature regime, and hardgapped at $T=0$. 
\end{abstract}
\begin{multicols}{2}
\narrowtext
\tighten
Phase diagrams of HighTcSuperconductors (HTS) have been reported which show a
spin glass phase in between the antiferromagnetic and superconducting
ones \cite{chou,scalapino,wbrenig,mydosh}. 
Strontium doped HTS are the most prominent examples, 
where antiferromagnetism gives way to clear signatures of spin
glass (SG) like behaviour before superconductivity sets in at higher
(hole)-doping. A typical SG order parameter was
identified at moderately low temperatures $T<8K$ \cite{chou} and even an 
infiltration within the superconducting domain at lowest temperatures was
described \cite{scalapino}. Classes of HTS exist as well,  
which do not seem to show spin glass and/or intermediate phase, or it
might have been impossible to detect it up to now.
Existence and nature of intermediate spin glass or
SG-alike phases in a certain doping range must be expected to be
important features of strongly correlated systems. Experiments
revealing close relations between magnetism and superconductivity in heavy
fermion (HF) systems addressed coexistence, phase separation, and
pairbreaking of local pairs by frozen moments for example
\cite{mydosh,steglich}. 
\\
Most of recent theories for HTS materials focussed on the
destruction of antiferromagnetism under doping. Viewing an
intermediate phase from the superconducting side, the appealing
concept of a nodal liquid \cite{MPAFisher} was developed, leaving aside
the notion and role of quenched disorder and randomness. 
Since we wish to deal here with superconductivity transitions 
under participation of a spin glass, a disorder model is a natural choice. 
Arguments on an important role of disorder in HTS and in HF-systems 
were provided experimentally and by theoretical reasoning 
\cite{mydosh,aharony,steglich}. 
Aspects such as non Fermi liquid behaviour, seen to arise in the 
vicinity of spin glass order \cite{georges}, also support this point of view. 
\\
In this Letter we present detailed results for spin glass to superconductor
transitions in a single-species fermionic model, which treats frustrated
magnetic-- and attractive interaction on the same footing. 
Unique features of the phase diagram are derived analytically and numerically.
We shall observe that for certain interaction ratios the location of the spin 
glass bears resemblance to that of a logarithmic resistivity regime residing 
above a spin glass ordered phase at lower $T$ in Sr-doped HTS \cite{aharony}.
A fluctuational state of broken down spin glass order 
should contribute to transport properties seen in intermediate phases above 
$T_f$. In particular, SG-order was recently shown to affect transport
properties strongly\cite{robrprb}, an effect that due to the random interaction
can well have a weak localization precursor.\\ 
We consider here a model described by the Hamiltonian
${\cal{H}}\equiv{\cal{H}}_{J v}+{\cal{H}}_t$, where
${\cal{H}}_{J v}=-\frac{1}{2}\sum J_{ij}\sigma_i\sigma_j-\sum
v_{ij}a_{i\downarrow}^{\dagger}a_{i\uparrow}^{\dagger}
a_{j\uparrow}a_{j\downarrow}-\mu\sum n_i$,
${\cal{H}}_t=\sum_{ij\sigma}t_{ij}a_{i\sigma}^{\dagger}a_{j\sigma}$,  
and $\sigma=n_{\uparrow}-n_{\downarrow}$,
$a_{i\sigma}$, and $n=n_{\uparrow}+n_{\downarrow}$ denoting
spin-, fermion-, and fermion-number operators respectively.
The variance $J^2$ of the frustrated and Gaussian--distributed magnetic 
interaction $J_{ij}$ and its magnitude relative to that of the attractive
coupling, $v_{q=0}/J$, are relevant parameters of our analysis below, 
together with the chemical potential and the related filling factor
$\nu(\mu)$.
We restrict the discussion to the small $t/J$ regime, which implies that
the selfconsistently determined magnetic band is much larger than the
hopping bandwidth in general. Only deep within the superconducting regime, 
where the magnetic bandwidth almost shrank to zero, the single fermion
hopping bandwidth becomes dominant.
Local pairs can be delocalized due to finite range $v_{ij}$ and/or by
arbitrarily weak fermion hopping $t_{ij}$. We employ a local
pairing theory of superconductivity based on the order parameter
$\Delta\equiv\langle a_{i\uparrow}a_{i\downarrow}\rangle$ and on a
two-particle coherence length replacing the usual one-particle 
length in the corresponding Ginzburg-Landau theory. 
One-fermion Green's functions \cite{oppermann} are local as in the 
$d=\infty$ method \cite{vollhardt,GeKr}. Superconductivity is of 
Bose condensation type 
and crosses over into BCS type behaviour (not considered here) in the
limit of small ratios between $v$ and hopping bandwidth. 
\\
The phase diagram illustrated in Figures 1 and 2 is obtained from the
free energy for model ${\cal{H}}_{Jv}$ given by
\begin{eqnarray}
f&=&\frac{1}{4}\beta
J^2((\tilde{q}-1)^2-(q-1)^2)+\frac{|\Delta|^2}{v}\nonumber\\
& &\hspace{1cm}-T\int_{-\infty}^{\infty}\frac{dz}{\sqrt{2\pi}}e^{-z^2/2}
\ln{\cal{C}},
\end{eqnarray}
where ${\cal{C}}\equiv \cosh(\beta \tilde{H})+
\cosh(\beta\sqrt{|\Delta|^2+\mu^2})\exp(-\frac{1}{2}\beta J^2\chi)$ 
with
$\chi=\beta(\tilde{q}-q)$ and effective field $\tilde{H}\equiv J\sqrt{q}z$; 
the saddle point solutions
$q$, $\tilde{q}$, and $\Delta$ are found from the coupled
selfconsistent equations obtained by extremizing f. 
The physical solutions of these coupled equations are not easy to identify, 
even outside the SG-phase; multiple solutions exist and changeovers of 
stability occur as chemical
potential $\mu$, temperature T, or interaction ratio $v/J$ are varied.
This leads to the complexity of the phase diagram displayed in 
Figures 1 and 2. 
\begin{figure}
\epsfxsize=8cm
\epsfbox{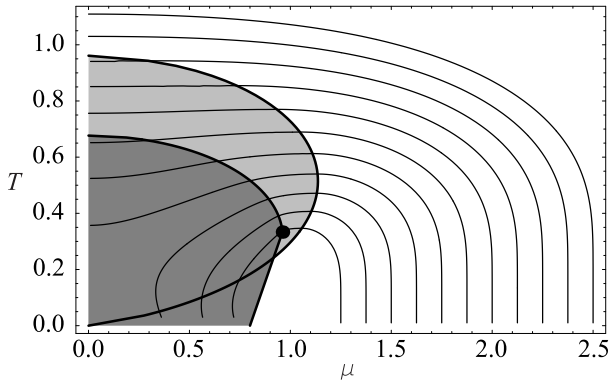}
\caption{Phase diagram of competing spin glass (SG) and superconducting 
(SC)-transitions (thin lines) for various attractive couplings $v$ 
(given by $v=2J\mu_{e.p.}$ at lines' right endpoint
($T=0,\mu=\mu_{e.p.}$)), at fixed $J=1$, and as a function of
chemical potential $\mu$ (dark grey area: maximum SG-domain for $v=0$).
Thin lines enclose SC-phase which remove pieces of the
($v=0$) SG-phase or suppress it totally at large enough $v/J$.
The bold line $l_3$ delimiting the light grey area joins
tricritical points ($T_c$-maxima) on SC-critical curves locating 1st order
transitions left of $l_3$
}
\end{figure}
Figures 1 and 2 reveal unique features of the competition between SG- and
SC-order. For example, an enhanced fermion concentration $\nu(\mu)$
reduces the effective spin density at larger $|\mu|$ and is seen to 
suppress the spin glass stronger than it eliminates the
superconducting phase as $\nu(\mu)\rightarrow$ 0 or 2. 
Within a large region above the SG-phase, 
the SC-critical curves become deformed, as Fig.1 shows for $v/J<4.5$, 
due to the increasing spin glass fluctuations feeded by the random
magnetic interaction. 
As the critical SC-curve passes through a maximum and starts to descend
with decreasing $\mu$, the SC-transitions change from second to first order.  
For still smaller ratios $v/J$ the SC-line enters the $v=0$ SG-phase:
in this case, magnetic moments freeze first and a discontinuous SG-SC
transition follows at lower temperature (shown for $v/J=3.25, 3.5, 3.75$).
For still smaller $v/J$ the 1st order SC-line falls rapidly to zero and
the SG-phase prevents superconducting order up to a critical value $\mu_c$
(see $v/J=3, 2.75,2.5$). The replica symmetric solution
also reveals the possibility of reentrant SC to SG-transitions, shown for
$v/J=2.5, 2.75, 3$ in Figure 1.\\
\begin{figure} 
\epsfxsize=8cm
\epsfbox{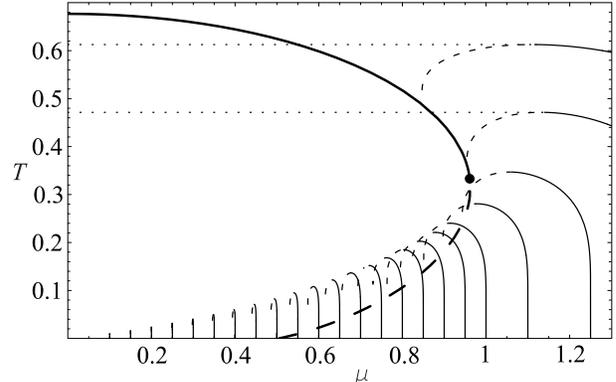}
\caption{Phase diagram close to the spin glass transition (bold line:
pure SG-transition, bold-dashed line: paramagnetic stability limit)
showing the stability limits of the superconducting (SC) phase (dotted
horizontal lines, given only for $v/J=3.5, 3.0$) and of the non-SC phase
(dashed thin lines) bifurcating at SC-tricritical points.
SC-critical (continued thin) lines are shown down to small $v/J$.}
\end{figure}
Figure 2 complements Figure 1 by adding, within a magnification of the
SG-SC boundary, the stability limits of the superconductor and
displays second order SC-transition lines 
crawling into the phase separation regime of the Ising spin glass
for small $v/J$ and $T$. For $T\rightarrow0$, the
quantum Parisi phase must be expected, which means that the
paramagnetic stability limit should be shifted to the left. Lacking
the full Parisi solution for $\mu>0$ we employ numerical solutions 
of fermionic TAP-equations \cite{TAP,rehker} to arrive at a straight
line estimate of the discontinuous paramagnet-SG transition curve. 
Discontinuous SC-SG transitions for $\frac{v}{J}<2.5$ 
must occur in between this curve and the SC-SG transition at
$\frac{v}{J}=2.5$. It is currently not possible 
to obtain a precise position of these discontinuous transitions
at low T, since this would require to unite the dynamic mean field theory
with the Parisi solution at infinite replica symmetry breaking (RSB)
(despite the discontinuity, one step RSB is insufficient at and near
$T=0$). An estimate can be obtained from the few percent rise of the
SG-energy due to RSB \cite{parisi}.
Furthermore, we arrive at the following conclusions:\\
i) There is no coexistence of spin glass - and local pairing 
superconducting order parameter in zero magnetic field. The detailed
analysis of the free energy and of all stability conditions shows only
SC-SG transitions between $\{q\neq0,\Delta=0\}$ and $\{q=0,\Delta\neq0\}$
for $H=0$. \\ 
ii) The transition between the two phases is always discontinuous and
exists only within a certain range of chemical potentials 
(or filling).\\
iii) For large enough $v/J$, like $v/J>4.13$ at half-filling for
example,the spin glass is prevented by the superconducting transition,
which is continuous for $v/J>4.55$ at half-filling; 
below this value it becomes discontinuous. 
The tricritical line, which separates these domains is included in Figure
1 for several values of $v/J$ and as a function of the chemical
potential.\\
iv) Decreasing the temperature at fixed $\mu<.96$ (see Figure 1), 
a second transition from SG to superconductivity occurs at
$T_c<T_{f}(v=0)$. 
Figures 1 and 2 display the exact numerical results of the
replica symmetric theory. It is known that the spin glass free energy
increases a few percent as an effect of replica symmetry breaking, 
which should slightly enlarge the superconducting domain 
(these corrections increase with decreasing temperature). 
Recalling the change of the spin glass-ferromagnetic boundary 
under RSB \cite{binderyoung} the reentrant
behaviour from superconductor to spin glass and back to superconductor
may disappear as well, but the SC-SG boundary will not become a vertical
line (unlike the boundaries between SG and ferro- or antiferromagnetic
phase in a purely magnetic model \cite{binderyoung};

\begin{figure}
\epsfxsize=8cm
\epsfbox{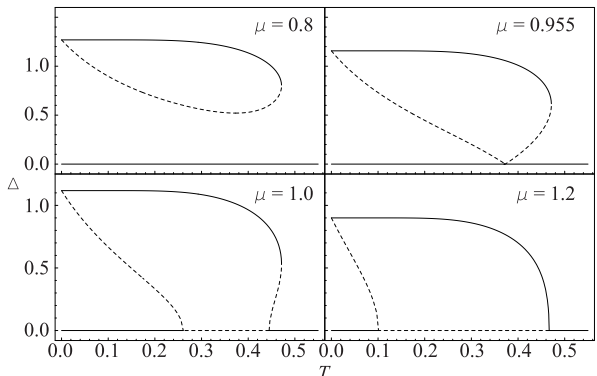}
\caption{Selfconsistent solutions for the superconducting order parameter
$\Delta$ (continued lines), shown for $v/J=3,
q=0$ and various $\mu$ values. Unstable solutions (dashed lines,
maxima of f) explain stability limits and different
discontinuous scenarios arising for sufficiently small $\mu$.} 
\end{figure}
Figure 3 shows the effect of random magnetic fluctuations, described by 
$\tilde{q}(T,\mu)$, on the superconducting order parameter for
exemplary values of $\mu$ and $v/J=3$.
The absence of coexistent order parameters allows to
set $q=0$. For $\mu = 0.8$ a first order transition from $\Delta=0$ to SC
order occurs at a temperature below the tricritical $T_{3}=0.475 J$,
the latter being independent of $\mu$. For $\mu=J$, the first order 
transition still occurs, but in addition the non-SC phase becomes 
locally unstable between the two zeros of the finite $\Delta$ solutions. 
At low temperatures, $\Delta=0$ is locally stable again, but the 
global minimum remains at $\Delta \neq 0$. Beyond the tricritical point, 
$\mu>\mu_3(\frac{v}{J}=3)=1.14 J$ in Figure 3, the stable 
SC solution continuously decreases to $\Delta=0$.\\ 
In a magnetic field new aspects arise: the transition temperature of the
superconductor will be reduced and finally vanish for $H> H_{c2}$,
leaving a smeared spin glass transition for sufficiently small
chemical potential. The overlap parameter $q$ is nonzero in a field and
then coexists with $\Delta$, since the field can penetrate the present
type II superconductor partially. In this case the Almeida Thouless line
can enter the superconductor, infiltrating ergodicity breaking there.\\
We analyze the superconducting phase by means of
the Green's functions: we derive the normal one
${\cal{G}}(\epsilon_n)$ , the anomalous one ${\cal{F}}(\epsilon_n)$,
and relevant particle-hole-$\Pi_{ph}(\omega_m)$ and particle-particle
bubble diagrams $\Pi_{pp}(\omega_m)$ for various limits. Superconductivity
arising in the magnetic band much larger than the hopping band is
described by the (local) anomalous Green's function
\begin{eqnarray}
& &{\cal{F}}(\epsilon)=\frac{i\hspace{.1cm}\Delta\hspace{.1cm}
\exp(\beta^2\tilde{q}/2)/(8\tilde{q}r_1)}{\cosh(\beta
r_1)+\exp(\beta^2\tilde{q}/2)}\\
& &\sum_{\{\lambda,s\}=\pm 1}(\epsilon+i\lambda r_1+i
s\beta\tilde{q})\hspace{.1cm}{\cal{U}}\hspace{-.1cm}
\left[1,\frac{3}{2},\frac{(\epsilon+i\lambda
r_1+i
s\beta\tilde{q})^2}{2\tilde{q}}\right]+\nonumber\\
& &\frac{i\hspace{.1cm}\Delta\hspace{.1cm} \cosh(\beta r_1)/(4\tilde{q}
r_1)}{\cosh(\beta
r_1)+\exp(\beta^2\tilde{q}/2)}\sum_{\lambda\pm
1}(\epsilon+i\lambda
r_1)\hspace{.1cm}{\cal{U}}\hspace{-.1cm}\left[1,\frac{3}{2},\frac{(\epsilon
+i r_1)^2}{2\tilde{q}}\right]\nonumber
\end{eqnarray}
where $r_1\equiv\sqrt{\mu^2+\Delta^2}$, $\epsilon\equiv(2n+1)\pi T$,
$J=1$, ${\cal{U}}$ denotes the hypergeometric U-function (Kummer
function) \cite{AbraStegun}, and the SC order parameter obeys
$\Delta=v T\sum_{\epsilon}{\cal{F}}(\epsilon)$.  
\begin{figure}
\vspace{-4.5cm}\hspace{-.3cm}
\psfig{file=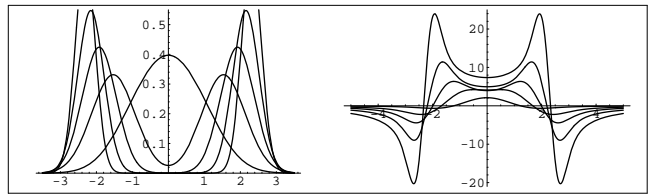,width=8.5cm,angle=0}
\vspace{-4.5cm}
\caption{Density of states $\rho(E)$ (left) and ${\cal{R}}e\{{\cal{F}}(E)\}$ 
(right), for ($v=5J,\mu=0,T_c\approx 1.11J$) and real energies E, 
show crossover from strongly to weakly gapless superconductivity
for decreasing temperatures $T/J=1.1, 1.0, .9, .8$ and $.7$}
\label{Frho}
\end{figure}
The solution for ${\cal{G}}(\epsilon)$, being similarly given in terms of 
the Kummer function, yields the density of states plotted together with 
${\cal{R}}e\{{\cal{F}}\}$ in Figure \ref{Frho} (weak fermion hopping- 
and dynamic corrections from $\Pi_{ph}(\omega)$ and $\chi(\omega)$ are 
negligible here). These plots of course employ the stable selfconsistent 
solutions for $\tilde{q}(T,\mu)$ and $\Delta(T,\mu)$ inserted in (2). 
They demonstrate the crossover from strongly gapless superconductivity
just below $T_c$ to a pronounced pseudogap.
Although invisibly small for temperatures lower than roughly 20\% below
$T_c$, the density of states remains nonzero in the pseudogap regime and
vanishes there only at $T=0$. The term gapless superconductor should
be used with care, since over a wide temperature range 
the gap looks almost perfect. The rounding of the superconducting gap
edges and the retarded way it opens up below $T_c$ is reminiscent of the
magnetic gap found below spin glass transitions \cite{robrprb}. As the 
superconducting gap sharpens at lower $T$, the magnetic band narrows 
and eventually the fermion hopping (bandwidth) starts to dominate. 
Then the anomalous Green's function crosses over into the hopping band 
solution which we derived for gaussian random $t_{ij}$ as
\begin{equation}
{\cal{F}}(\epsilon)=-\frac{\Delta}{2
w^2}+\frac{\Delta}{\sqrt{2}w^2}
\frac{\sqrt{w^2+\epsilon^2+\Delta^2-\mu^2+r(\epsilon)}}{\sqrt{\epsilon^2+
\Delta^2}}, 
\label{f6}
\end{equation}
where $r(\epsilon)=\sqrt{4 w^2(\epsilon^2+\Delta^2)+(-w^2+\epsilon^2+
\Delta^2+\mu^2)^2}$ with $w^2\equiv 4\langle t^2_{ij}\rangle$.
Equation (\ref{f6}) is an exact local solution which holds for arbitrary
bandwidth $w$ and chemical potential $\mu$, extending an earlier
result for half-filling \cite{oppermann}. 
The transition temperature derived from (\ref{f6}) illustrates in
Figure \ref{crossover} the crossover from linear Bose
condensation $T_c(v)$-dependence to exponential BCS-type.
\begin{figure}
\epsfxsize=8.5cm
\epsfbox{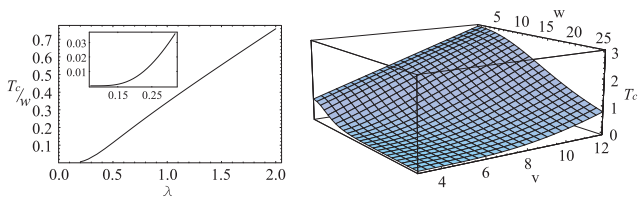}
\caption{Complete crossover from Bose condensation type to BCS behaviour
illustrated in 3D plot of $T_c(v)$ as $w/v$ increases; typical values for 
both linear and exponential dependence are given as a function of
$\lambda\equiv v\rho_F$} 
\label{crossover}
\end{figure}
The Bose-BCS crossover can be followed through the whole parameter range 
due to the local property of the one-particle Green's functions, a feature
appreciated as well in the $d=\infty$ method for clean systems
\cite{vollhardt,GeKr}. 
The suppression of one particle phase coherence, due
to infinite dimensions in clean systems or by symmetry
requirement in the quenched average used here, results in type II
superconductivity with two particle coherence length replacing the
standard one in the Ginzburg Landau theory (many details including 
paramagnetic pairbreaking were published in \cite{oppermann})
The puzzling coexistence phenomena in zero field, absence of 
$(q\neq0,\Delta\neq0)$ solutions but presence of phase separation regimes, 
and the related discontinuous transition between the magnetic and
superconducting phases is
further elucidated by the bubble diagram $\Pi_{pp}(T,\omega)$.
At half-filling for example, the condition $1=v\Pi_{pp}(T_c,\omega=0)$
yields $T_c=(1-\tilde{q}(T_c))\frac{v}{4}$, ($\tilde{q}=T
\chi|_{_{\omega=0}}$).
Equating this with the spin glass freezing temperature
$T_f=J\tilde{q}(T_f)$,
ie solving formally $T_c=T_f$ to search for simultaneous onset of both
types of order, results in $\tilde{q}=v/(v+J)$. Since
$\tilde{q}(T_f)=.6767$ one finds $v=4.1862$ (let $J=1$). 
Since $\max\{\Pi(T,\mu=0,\omega=0)\}=.242$ at $T=.731$ no solution exists
for $v<v_{\min}=4.14$, while two solutions are found for
$v=4.186>v_{\min}$,
only those with $T>.731$ being free energy minima.
At half-filling, the thermal first order superconducting transition occurs at
temperature $T_{c1}>T_f$ for $v>3.8$, but the stability limit of the normal 
phase is absent for $3.8<v<4.14$; spin glass order will not reappear down to
lowest temperatures. For $v<3.8$ the spin glass transition occurs at 
$T_f=.6767$ with a first order SG-SC transition to follow at 
lower temperature.\\  
For the sake of transparency we discussed only fully frustrated magnetic
interactions and zero field phenomena. Subsequent antiferromagnetic-spin
glass-superconductor transitions, as $\mu$ increases from zero are not
simply obtained once the $J_{ij}$ is given an antiferromagnetic
part. 
Thus, models extended by the Hubbard interaction, but also the allowance for 
different symmetries of the order parameter, and the dynamic quantum Parisi 
phase
are examples for future research on links between antiferromagnetism, spin
glass order, and superconductivity.\\ 
The model we analyzed here proved to have a special type of gapless
superconductivity due to the vicinity of spin glass order and due to
correlations induced by the frustrated magnetic interaction.\\
We acknowledge support by the DFG under grant Op28/5-1 and by the SFB410.

\end{multicols}
\end{document}